\documentclass{aa}
\usepackage{graphicx}
\usepackage{natbib}
\bibpunct{(}{)}{;}{a}{}{,}

\def\aap{A\&A}
\def\aaps{A\&AS}
\def\aj{AJ}
\def\apj{ApJ}
\def\apjs{ApJS}

\def\apjl{ApJ}

\def\mnras{MNRAS}

\def\pasp{PASP}
\def\pasj{PASJ}

\begin{document}

\title{Early-type stars in the young open cluster IC\,1805} 
\subtitle{II. The probably single stars HD\,15570 and HD\,15629, and the massive binary/triple system HD\,15558\thanks{Based on observations collected at the Observatoire de Haute-Provence (France).}}

\author{M. De Becker
	\inst{1} 
	\and
	G. Rauw
	\inst{1}\fnmsep\thanks{Research Associate FNRS (Belgium).}
	\and
	J. Manfroid
	\inst{1}\fnmsep\thanks{Research Director FNRS (Belgium).}
	\and 
	P. Eenens
	\inst{2}
	}

\offprints{M. De Becker}

\institute{Institut d'Astrophysique et de G\'eophysique, Universit\'e de Li\`ege, 17, All\'ee du 6 Ao\^ut, B5c, B-4000 Sart Tilman, Belgium\\
\and
Departamento de Astronomia, Universidad de Guanajuato, Apartado 144, 36000 Guanajuato, GTO, Mexico}

\date{Received ; accepted }

\abstract
{}
{We address the issue of the multiplicity of the three brightest early-type stars of the young open cluster IC\,1805, namely HD\,15570, HD\,15629 and HD\,15558.}
{For the three stars, we measured the radial velocity by fitting Gaussian curves to line profiles in the optical domain. In the case of the massive binary HD\,15558, we also used a spectral disentangling method to separate the spectra of the primary and of the secondary in order to derive the radial velocities of the two components. These measurements were used to compute orbital solutions for HD\,15558.}
{For HD\,15570 and HD\,15629, the radial velocities do not present any significant trend attributable to a binary motion on time scales of a few days, nor from one year to the next. In the case of HD\,15558 we obtained an improved SB1 orbital solution with a period of about 442 days, and we report for the first time on the detection of the spectral signature of its secondary star. We derive spectral types O5.5III(f) and O7V for the primary and the secondary of HD\,15558. We tentatively compute a first SB2 orbital solution although the radial velocities from the secondary star should be considered with caution. The mass ratio is rather high, i.e. about 3, and leads to very extreme minimum masses, in particular for the primary object. Minimum masses of the order of 150\,$\pm$\,50 and 50\,$\pm$\,15\,M$_\odot$ are found respectively for the primary and the secondary.}
{We propose that HD\,15558 could be a triple system. This scenario could help to reconcile the very large minimum mass derived for the primary object with its spectral type. In addition, considering new and previously published results, we find that the binary frequency among O-stars in IC\,1805 has a lower limit of 20\,\%, and that previously published values (80\,\%) are probably overestimated.}


\maketitle

\section{Introduction} \label{intro}

The study of stellar populations of young open clusters has been the purpose of many works in the last years (e.g. \citealt{Sag}; \citealt{RM}). One of the crucial questions addressed in this context concerns the massive star content and the binary frequency of the earliest stars harboured by these clusters. Recent numerical simulations of the star formation in open clusters predict a high level of mass segregation, along with the formation of the most massive stars through simultaneous accretion and stellar collisions, resulting in the presence of the most massive stars in binary systems \citep{BB2002}. The investigation of the multiplicity of the massive star population of young open clusters is consequently of particular interest. For instance, \citet{GM1805} presented radial velocity measurements for 37 O- and B-stars in the open cluster NGC\,6231, and proposed binary frequencies for a series of open clusters, among which is IC\,1805.
 
IC\,1805 is a young rich open cluster in the core of the Cas\,OB6 association, in the molecular cloud W\,4 in the Perseus spiral arm of our Galaxy. \citet{massey1805} inferred an age of 1-3\,Myr for the cluster, in agreement with other previous estimates (see \citealt{fein1805}, and references therein, although these latter authors proposed an age $<$\,1\,Myr). The spectroscopic study of \citet{SH1805} revealed that about 80 of the members of IC\,1805 are O or B stars. \citet{GM1805} (see also \citealt{Ish}) estimated a binary frequency of 80\,\% among the 10 O-stars in IC\,1805. However, in a previous paper \citep[][Paper I]{ic1805_1}, we showed that two suspected binaries, BD\,+60$^{\circ}$\,501 (O7V((f))) and BD\,+60$^{\circ}$\,513 (O7.5V((f))), were most probably single stars, suggesting that the binary frequency proposed by \citet{GM1805} might be overestimated. In this paper, we investigate the multiplicity of the three earliest O-type stars of the cluster: HD\,15570, HD\,15629, and HD\,15558.\\

HD\,15570 (O4If$^+$, V\,=\,8.10) was proposed to be the most massive member of IC\,1805, and incidentally one of the most massive and most luminous stars known in our Galaxy, with a present evolutionary mass of about 80\,M$_\odot$ following \citet{HPV}. HD\,15629 (O5V((f)), V\,=\,8.42) has also been proposed to be a very massive star. According to \citet{HPV}, this star could have evolved from an initial mass of about 70\,M$_\odot$, and presently be on the way towards evolutionary stages close to those of HD\,14947 (O5If$^+$) and HD\,210839 (O6I(n)fp, $\lambda$\,Cep). The same authors inferred a present-day evolutionary mass of about 61\,M$_\odot$. For both stars, no clear evidence of binarity has been reported in the literature, even though some radial velocities quoted in the WEBDA data base\footnote{Available at http://obswww.unige.ch/webda} suggest the occurrence of variations.

HD\,15558 (m$_{V}$\,=\,8.0) was classified as an O5III(f) star by \citet{Mat2}. It was reported to be a spectroscopic binary with a period of about 420\,d for the first time by Trumpler (according to \citealt{und67}). Up to now, the only orbital solution proposed for this system is that of \citet{GM}, revealing a highly eccentric binary (e\,=\,0.54) with a period of about 440\,d, i.e. the longest of any O-type spectroscopic binary known at that time. This star is also believed to be very massive. \citet{HPV} derived spectroscopic and evolutionary masses of about 90\,M$_\odot$ for HD\,15558. Finally, this massive binary is classified as a non-thermal radio emitter. In this context, the study of its binarity is of crucial interest \citep[see][ for a detailed discussion]{thesis}.\\

This paper is organized as follows. Section\,\ref{spec} presents the optical spectrum of the stars. The results of the radial velocity study for HD\,15570 and HD\,15629 are given in Sect.\,\ref{single}. The case of HD\,15558 is discussed in detail in Sect.\,\ref{hd15558}. Section\,\ref{disc} consists of a discussion of our results. A summary of the main results and the conclusions are provided in Sect.\,\ref{conc}.

\section{The optical spectrum\label{spec}} 

\begin{figure*}
\centering
\includegraphics[width=120mm]{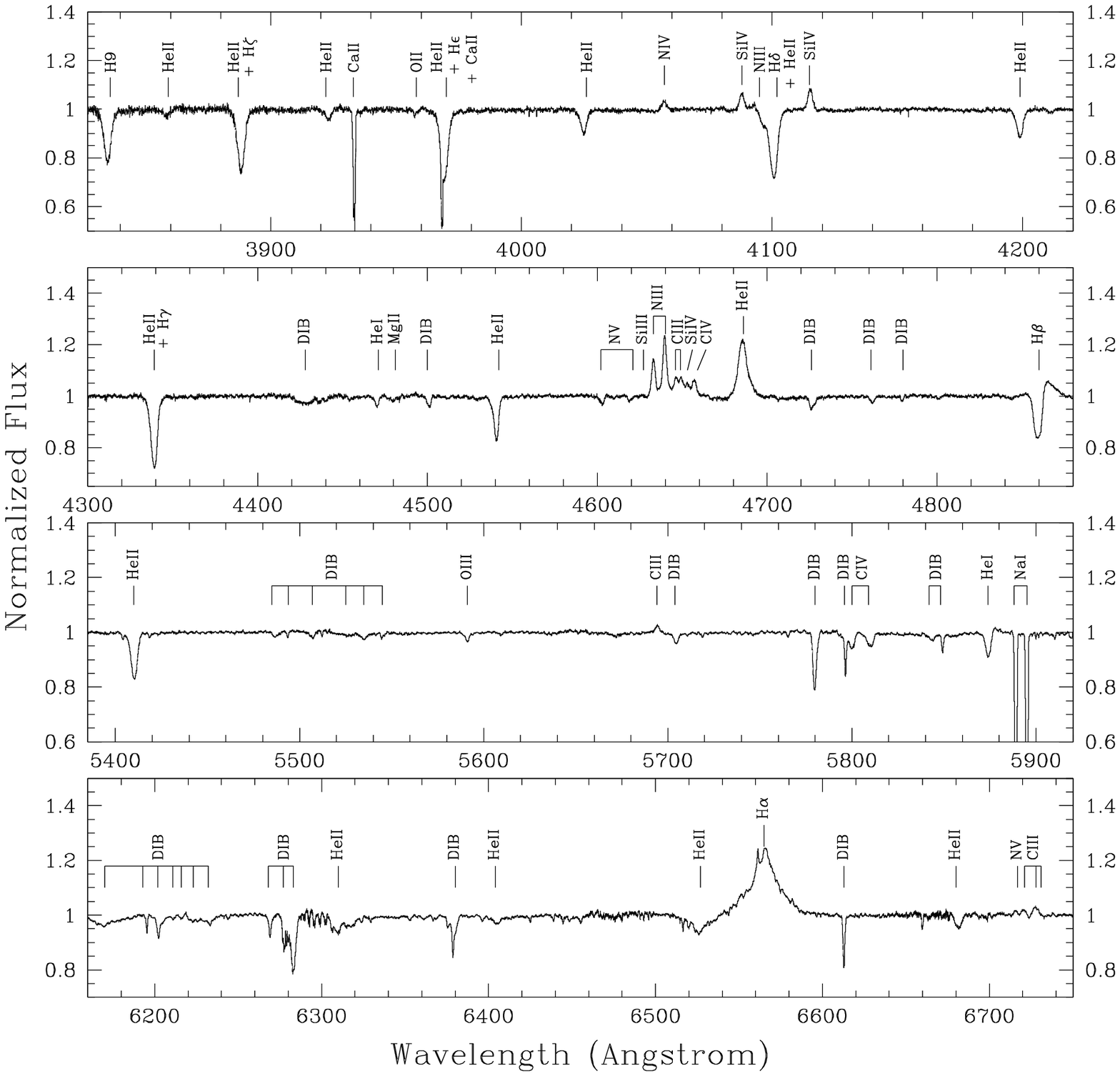}
\caption{Mean spectrum of HD\,15570 calculated from spectra obtained at SPM between HJD 2\,453\,287.93 and HJD 2\,453\,290.00.}
\label{spm}
\end{figure*}

\begin{figure*}
\centering
\includegraphics[width=120mm]{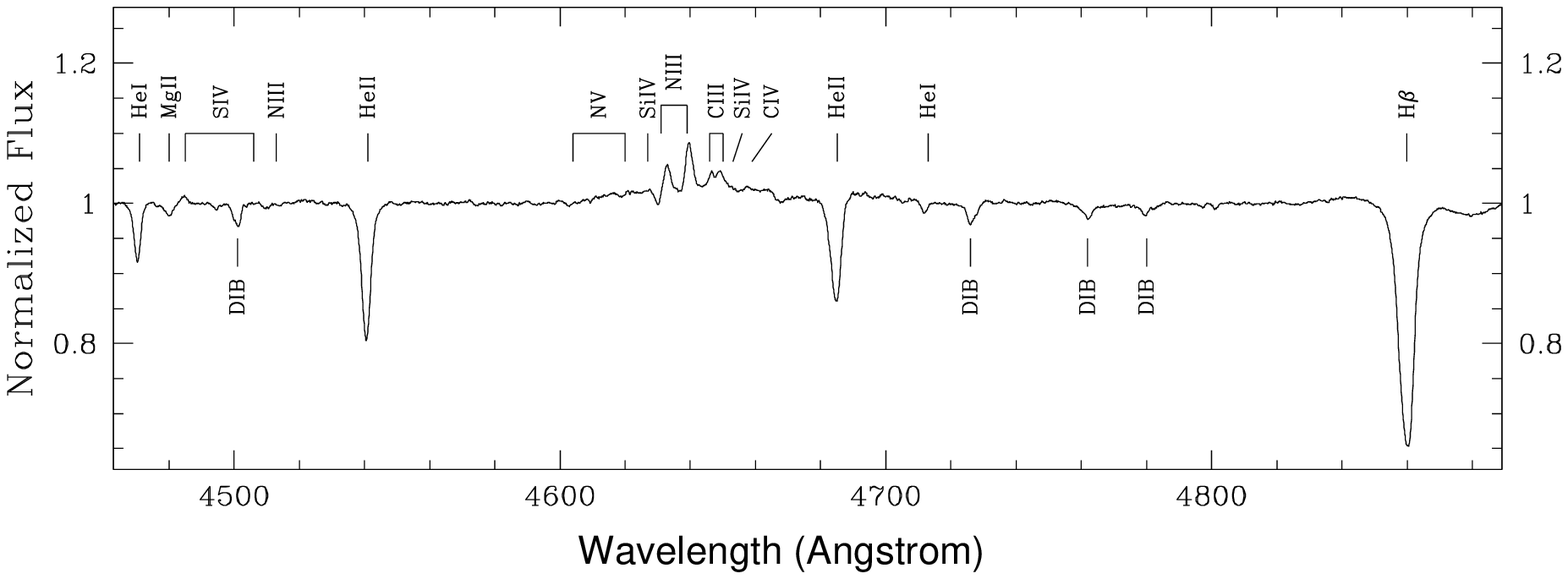}
\caption{Mean normalized spectrum of HD\,15629 between 4465 \AA\, and 4890 \AA\, as observed in September 2002 at OHP.}
\label{ohp}
\end{figure*}

\begin{figure*}
\centering
\includegraphics[width=120mm]{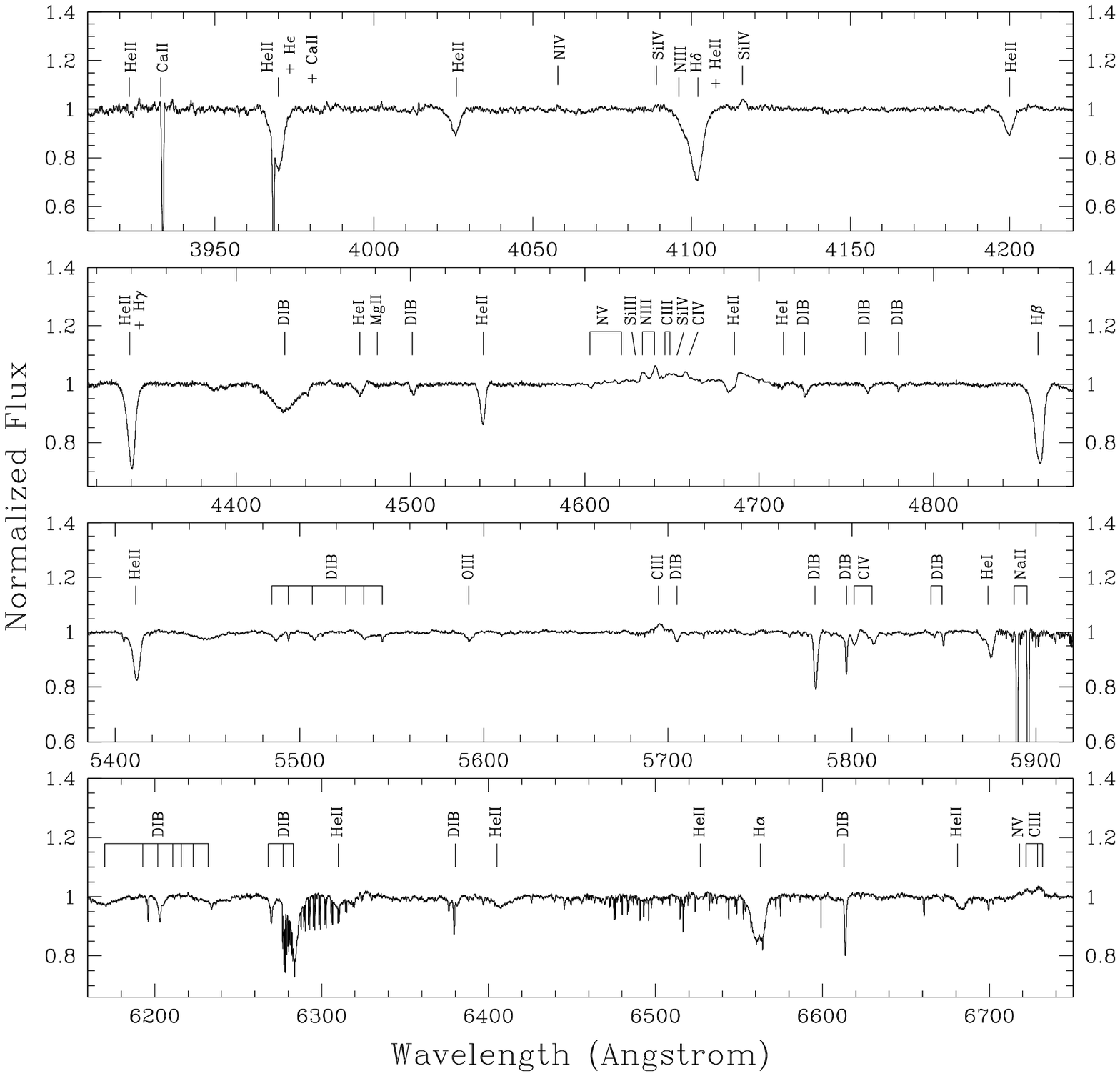}
\caption{Spectrum of HD\,15558 obtained about at phase 0.8, i.e. at the maximum of the radial velocity curve presented in Fig.\,\ref{orbsol}. This spectrum was obtained with the Elodie echelle spectrograph on 7 January 2004, except for the spectral domain between 4580 and 4710\,\AA\, which is taken from one of the spectra obtained in September 2001 with the Aur\'elie spectrograph. We note that a large portion of the spectrum was skipped because of a lack of important spectral features.\label{ohp2}}
\end{figure*}

\subsection{HD\,15570}
A detailed description of our data is given in Appendix\,\ref{obs}. The most prominent features observed in the spectrum presented in Fig.\,\ref{spm} are the \ion{He}{ii} $\lambda$ 4686 and H$\alpha$ lines, in strong emission. The \ion{He}{ii} $\lambda$ 4686 emission is one of the characteristics of the Of$^+$ type, along with \ion{Si}{iv} $\lambda$$\lambda$ 4088-4115 also in emission and \ion{N}{iii} $\lambda$$\lambda$ 4634-4641. We find also weaker emission lines of \ion{N}{iv} at 4058\,\AA\,, \ion{C}{iii} at 4650 and 5696\,\AA\,, and also \ion{C}{iv} $\lambda$ 4662. The weak emission lines at 6721, 6728 and 6731\,\AA\, might be attributed to \ion{C}{iii} \citep{wal2001}. We report also on the probable presence of \ion{Si}{iv} $\lambda$ 4656 and \ion{N}{v} $\lambda$$\lambda$ 6716,6718 in emission \citep{wal2001}. The H$\beta$ line appears as a P\,Cygni profile with a very deep absorption component. Other absorption lines are mainly due to \ion{He}{ii}. \ion{He}{i} $\lambda$ 4471 is very weak. We also note that the \ion{He}{i} $\lambda$ 5875 line appears as a P\,Cygni profile, even though the emission component is rather weak. The \ion{Mg}{ii} $\lambda$ 4481 line is detected, and we also note the presence of the \ion{N}{v} $\lambda$$\lambda$ 4604,4620 lines in absorption. Even though its presence is not obvious in the mean SPM spectrum (see Fig.\,\ref{spm}), a weak emission feature at about 4487 \AA\, is observed in OHP spectra and is probably due to \ion{S}{iv} \citep{WR}.

Using the classification criterion given by \cite{Mat}, we infer the spectral type O4. The various emission lines discussed above lead to the O4If$^+$ spectral type for HD\,15570, in agreement with Walborn's classification \citep{wal1972}. A comparison with the spectrum of the O4If$^+$ star HD\,269698 \citep{WF} lends further support to this classification.

\subsection{HD\,15629}
The strongest lines in the blue spectrum of HD\,15629 (Fig.\,\ref{ohp}) are in order of decreasing intensity: H$\beta$, \ion{He}{ii} $\lambda$ 4542, \ion{He}{ii} $\lambda$ 4686 and \ion{He}{i} $\lambda$ 4471, all in absorption. The \ion{He}{i} $\lambda$ 4713 absorption line is very weak. The \ion{Mg}{ii} $\lambda$ 4481 absorption line is also present. The \ion{N}{iii} $\lambda$$\lambda$ 4634-4641 lines are the strongest emission features observed in this spectrum. The \ion{C}{iii} blend at about 4650 is also present and we note the possible presence of \ion{C}{iv} $\lambda$ 4662 in weak emission. The weak emission feature at about 4487 \AA\, is probably due to \ion{S}{iv} \citep{WR}.

Adopting the classification criterion porposed by \citet{Mat}, we derive an O5 spectral type. Considering that the \ion{N}{iii} $\lambda$$\lambda$ 4634-4641 features are in emission, and that \ion{He}{ii} $\lambda$ 4686 is in strong absorption, we adopt the O5V((f)) spectral type for HD\,15629, in agreement with previous classifications.

\subsection{HD\,15558}

In Fig.\,\ref{ohp2}, we present the spectrum of HD\,15558 between 3910 and 6750\,\AA\, obtained with the Elodie spectrograph on 7 January 2004, i.e. close to the maximum of the radial velocity curve presented in Fig.\,\ref{orbsol}. We note that the orders of the echelle spectra obtained with Elodie are rather narrow, rendering the rectification procedure somewhat difficult where the wings of several lines merge over several tens of \AA\,. So, for the purpose of the presentation of the optical spectrum of HD\,15558, the spectral domain between 4580 and 4710\,\AA\, in Fig.\,\ref{ohp2} comes from one of the Aur\'elie spectra obtained during the September 2001 observing run, i.e. very close to the orbital phase of the selected Elodie spectrum.

The strongest absorption features are the hydrogen Balmer lines. The \ion{He}{ii} lines appear also in absorption, except for \ion{He}{ii} $\lambda$ 4686 which displays a P-Cygni profile. The \ion{He}{i} absorption lines at 4471 and 5875\,\AA\, are also present, along with \ion{He}{i} $\lambda$ 4713 which is very weak. The other absorption features observed in the spectrum are \ion{Mg}{ii} $\lambda$ 4481, \ion{N}{v} $\lambda$$\lambda$ 4604,4620, Si\,{\sc iii} $\lambda$ 4627, \ion{O}{iii} $\lambda$ 5592 and \ion{C}{iv} $\lambda$$\lambda$ 5801,5812. The strongest emission lines are the \ion{N}{iii} $\lambda$$\lambda$ 4634,4641 lines. We report also on noticeable \ion{C}{iii} lines at 4650, 4652, 5696, 6721,6728 and 6731\,\AA\,. The three latter lines were qualified as selective emission lines by \citet{wal2001} as well as the \ion{N}{v} $\lambda$$\lambda$ 6716,6718 emission features. We report also on the presence of the \ion{Si}{iv} $\lambda$ 4116 emission line, along with that of the much weaker \ion{Si}{iv} $\lambda$ 4088. Finally, the \ion{C}{iv} $\lambda$ 4662 emission line is cleary present. The spectral classification of the stars in this binary system is detailed in Sect.\,\ref{class58}.

\section{The probably single stars HD\,15570 and HD\,15629} \label{single}
Both stars have been mainly observed at the Observatoire de Haute-Provence (OHP, France) with the 1.52\,m telescope. Some spectra were also obtained with the 2.1\,m telescope at the Observatorio Astron\'omico Nacional of San Pedro Martir (SPM, Mexico). For details on the observations and the data reduction procedure, see Appendix\,\ref{obs}. We have measured the radial velocity (RV) of the strongest lines for HD\,15570 and HD\,15629 by fitting Gaussians to the line profiles. The RVs of various lines from OHP data only are quoted in Table\,\ref{rvtab}.

In the case of HD\,15570, the dispersion of the RV of the \ion{He}{i} $\lambda$ 4471 line is larger than for other lines in most of our data sets, as its profile deviates significantly from a Gaussian. In the case of the \ion{He}{ii} $\lambda$ 4542 line, we performed a Fourier analysis of the RVs following the method described by \citet{HMM} and revised by \citet{gosset30a}. The periodogram reveals several peaks. The highest one is found at a frequency of 0.71\,d$^{-1}$. However, the corresponding semi-amplitude is very low (about 4\,km\,s$^{-1}$) compared to the typical error on the RVs. This error, i.e. about 10\,km\,s$^{-1}$, corresponds to the standard deviation determined for the radial velocity of a Diffuse Interstellar Band (DIB) at about 4762 \AA\,. The radial velocities of the \ion{N}{iii} emission lines appear stable on the time scales investigated in this study. As a result, we do not detect any significant RV variation attributable to binarity on time scales of a few days, nor from one year to the next, in our OHP data. The radial velocities measured on the main absorption and emission lines observed in the SPM spectra did not reveal any significant RV change neither on a time scale of a few days, although we may note that the dispersion obtained in most cases is larger due to non-Gaussian profiles and to a poorer quality of the data. The lack of significant RV variations is not in agreement with the RV changes suggested by the data collected in the WEBDA data base. We caution however that these archive RV measurements were obtained on the basis of sometimes heterogeneous line lists and with instruments of different capabilities, at very different epochs.

The strong profile variability displayed by the \ion{He}{ii} $\lambda$ 4686 and H$\beta$ lines \citep[see][]{thesis}, along with their noticeable asymmetry, prevented us from obtaining accurate and reliable RVs from the entire profiles of these lines. Still, we mention that the RVs measured at the top (resp. bottom) of the \ion{He}{ii} $\lambda$ 4686 (resp. H$\beta$) line reveal rather strong wavelength shifts, which are correlated both in direction and amplitude (semi-amplitude of $\sim$\,30\,km\,s$^{-1}$) for the two lines. However, the lack of significant shift for the RVs of \ion{He}{i} $\lambda$ 4471, \ion{He}{ii} $\lambda$ 4542 and \ion{N}{iii} $\lambda$$\lambda$ 4634-4641 does not support the idea that this could be due to the orbital motion in a binary system. We finally note that the RVs of the Balmer lines measured on SPM spectra display the expected progression due to the transition from purely absorption (H9) to P-Cygni (H$\beta$) profiles.\\

For HD\,15629, we measured the RV for several lines that are free from profile variability. Therefore, only the \ion{He}{i} $\lambda$ 4471, \ion{He}{ii} $\lambda$ 4542, and \ion{N}{iii} $\lambda$$\lambda$ 4634-4641 lines are considered in this discussion. A Fourier analysis of the RVs from the three data sets reveals a highest peak at 0.08\,d$^{-1}$ (P\,=\,12.6\,d) for \ion{He}{i} $\lambda$ 4471, and at 0.16\,d$^{-1}$ (P\,=\,6.25\,d) for \ion{He}{ii} $\lambda$ 4542. The periodogram for this latter line presents also a peak close to that of the \ion{He}{i} line at 0.09\,d$^{-1}$ (P\,=\,11.1\,d). However, the amplitude of these peaks is only about 5\,km\,s$^{-1}$. We detect no significant RV variations for any of the lines. This result is in contrast with the variable status reported by \citet{und67} and \cite{Hum}. We also mention that long term RV variations were reported for this star (see the WEBDA data base), but we should consider these values with caution for the same reasons as for HD\,15570 (see above). In summary, we did not find any trend pointing to a binary scenario on the time scales sampled by our data. 

\section{The massive binary HD\,15558\label{hd15558}} 

\subsection{Radial velocity time series \label{rvts}}

As mentioned in Sect.\,\ref{intro}, the only orbital solution available for HD\,15558 was proposed by \citet{GM}. These authors reported on a period of 439.3 $\pm$ 1.0\,d, with an eccentricity of 0.54 $\pm$ 0.05. However, their orbital solution was based on time series including a somewhat heterogeneous set of radial velocities. They indeed used mean RVs calculated on the basis of 4 to 12 different lines. Among these lines, many are expected to be at least partly produced in the stellar wind, and might therefore not reflect the RV of the star itself. The heterogeneity of their RV time series could have a significant impact on the orbital parameters. In this section, we describe the procedure we adopted to establish our RV time series, and how we use it to determine the SB1 orbital parameters of the system.

The details on our data on HD\,15558 are given in Appendix\,\ref{obs}. First of all, we focused on the wavelength range covered by all our spectra, whatever the instrumentation used, i.e. the spectral domain between 4455 and 4680\,\AA\,. We then selected the lines whose profile did not deviate too strongly from a typical Gaussian shape. Only the \ion{He}{ii} $\lambda$ 4542 and the \ion{N}{iii} $\lambda$$\lambda$ 4634,4641 lines meet our criteria. We then determined the RVs by fitting Gaussians to the profiles. In the case of Elodie data, the \ion{He}{ii} line is simultaneously observed in two adjacent orders. We therefore measured the RV on each order and we used the mean of the two values in our time series. The RVs obtained from these three lines, along with those obtained from other lines observed only in our echelle spectra, are compiled in Table\,\ref{rvtab2}. In this Table, the column labeled `Mean' contains the mean RVs obtained from the three lines discussed above. We estimate that the expected uncertainty on the RVs are respectively of the order of 15-20, 10-15, and 5-8\,km\,s$^{-1}$ respectively for low resolution Aur\'elie, medium resolution Aur\'elie, and high resolution Elodie data. However, we mention that the error on the RVs measured on some low quality Elodie spectra may be significantly larger. These uncertainties were estimated on the basis of radial velocity measurements performed on DIBs.

\subsection{Period determination \label{per}}

We performed a Fourier analysis on our RV time series following the technique described by \citet{HMM} and revised by \citet{gosset30a}, as used for instance in Paper I and by \citet{Let8a} for other spectroscopic binaries. We independently applied the same technique to three RV time series obtained respectively from \ion{He}{ii} $\lambda$ 4542, a mean of the RVs from the two \ion{N}{iii} $\lambda$$\lambda$ 4634,4641 lines, and a mean of the \ion{He}{ii} and \ion{N}{iii} lines. In the three cases, the periodogram is dominated by a strong peak at a frequency of 0.00226\,d$^{-1}$, corresponding to a period of about 442\,d. As the time base covered by our data is about 1608\,d, the typical width of the peaks of the periodograms is about 6.22\,$\times$\,10$^{-4}$\,d$^{-1}$. Considering that the uncertainty on the frequency is about 10\,\% of the width of the peak, we obtain an uncertainty on the period of about 12\,d. The fact that the three RV time series lead exactly to the same period is a strong argument for this value being the true period of the system.

\subsection{Orbital solution \label{orb}}

We obtained the SB1 orbital solution of HD\,15558 using the method of \citet{WHS} for SB1 systems. We assigned different weights to take into account the expected uncertainties affecting our RV measurements. These weights vary between 0.1 and 1.0 respectively for very poor quality and good quality Elodie data. Aur\'elie RVs get intermediate weights to take into account the fact that though the resolution of the spectrograph is rather low, the quality of the data is much better than for Elodie spectra.

\begin{figure}
\centering
\includegraphics[width=80mm]{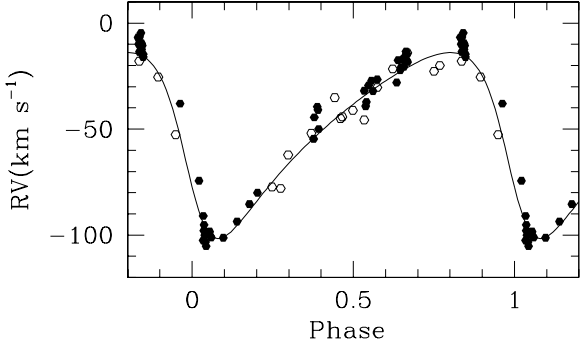}
\caption{Radial velocity curve of HD\,15558 for an orbital period of 442 d. The open and filled hexagons stand respectively for the primary RVs obtained with Elodie and Aur\'elie spectra. The solid line yields our best fit orbital solution respectively corresponding to the parameters provided by Table\,\ref{orbpar}. The mean of the RVs of the \ion{He}{ii} $\lambda$ 4542 and of the \ion{N}{iii} $\lambda$$\lambda$ 4634,4641 lines was used to compute this orbital solution. \label{orbsol}}
\end{figure}

\begin{table}
\caption{Orbital parameters of the SB1 solution of HD\,15558. T$_{\circ}$ refers to the time of periastron passage. $\gamma$, $K$, and $a\,\sin\,i$ denote respectively the systemic velocity, the amplitude of the radial velocity curve, and the projected separation between the centre of the star and the centre of mass of the binary system. The last row provides the mass function. The orbital elements are given for the solutions obtained respectively from the mean of the RVs of the \ion{He}{ii} $\lambda$ 4542 and \ion{N}{iii} $\lambda$$\lambda$ 4634,4641 lines, and from the RVs of \ion{He}{ii} $\lambda$ 4542 only. T$_\circ$ is expressed in HJD--2\,450\,000. \label{orbpar}}
\begin{center}
\begin{tabular}{l c c}
\hline\hline
\vspace*{-0.2cm}\\
 & \ion{He}{ii} $\lambda$ 4542 & \ion{He}{ii} $\lambda$ 4542 \\
 & \ion{N}{iii} $\lambda$$\lambda$ 4634,4641 &  \\
\vspace*{-0.2cm}\\
\hline
P (days)  & \multicolumn{2}{c}{442 (fixed)} \\
$e$   & 0.39 $\pm$ 0.03 & 0.40 $\pm$ 0.03 \\
T$_\circ$ & 1795.400 $\pm$ 6.556 & 1798.385 $\pm$ 7.207 \\
$\gamma$ (km\,s$^{-1}$)  & --50.2 $\pm$ 1.1 & --40.1 $\pm$ 1.1 \\
$K$ (km\,s$^{-1}$) & 43.9 $\pm$ 1.8 & 41.3 $\pm$ 1.8 \\
$\omega$ & 116$^\circ$ $\pm$ 6$^\circ$ & 121$^\circ$ $\pm$ 7$^\circ$ \\
$a\,\sin\,i$ (R$_\odot$) & 350.0 $\pm$ 15.2 & 331.2 $\pm$ 15.2 \\
$f(m)$ (M$_\odot$)  & 3.0 $\pm$ 0.4 & 2.5 $\pm$ 0.3 \\
\vspace*{-0.2cm}\\
\hline
\end{tabular}
\end{center}
\end{table}

We first fixed the period to the value determined from the Fourier analysis of our RV time series, i.e. $\sim$\,442\,d. We obtained similar results for the orbital parameters, i.e. for most of them within the error bars, whatever the RV time series used. We also calculated the orbital solution through an iterative process allowing the period to vary, but it did not improve significantly the results. We indeed obtained a period of about 445\,d, which can not be distinguished from the 442\,d period obtained in Sect\,\ref{per} provided the uncertainty on the period is about 12\,d. Therefore, we adopted the results obtained with a period fixed to 442\,d. The corresponding orbital parameters are quoted in Table\,\ref{orbpar} for the series of RVs obtained respectively on the basis of the \ion{He}{ii} and \ion{N}{iii} lines ({\it left column}), and on the basis of the \ion{He}{ii} line only {\it right column}.

\subsection{Searching for the companion \label{comp}}

We inspected more carefully spectra obtained at phases close to the extrema of the radial velocity curve presented in Fig.\,\ref{orbsol}. Though we did not obtain any echelle spectra close to the minimum of the primary radial velocity curve, the September 2000 and September 2001 Aur\'elie observing runs fall respectively very close to the minimum and maximum of the radial velocity curve. We were therefore able to perform a more careful inspection of the line profiles on the basis of our Aur\'elie spectra. We added together the 12 spectra obtained during each observing run to obtain two high signal-to-noise ratio spectra, respectively of about 1000 at the minimum and 900 at the maximum. We detected opposite asymmetries at both extrema for the profiles of the \ion{He}{i} $\lambda$ 4471, \ion{Mg}{ii} $\lambda$ 4481 and \ion{He}{ii} $\lambda$ 4542 lines, suggesting clearly the presence of the secondary. The inspection of the Elodie spectrum closest to the maximum of the primary RV curve reveals also the signature of the secondary in the profile of the \ion{C}{iv} $\lambda$ 5812 and \ion{He}{i} $\lambda$ 5876 lines.

\begin{figure}[ht]
\centering
\includegraphics[width=80mm]{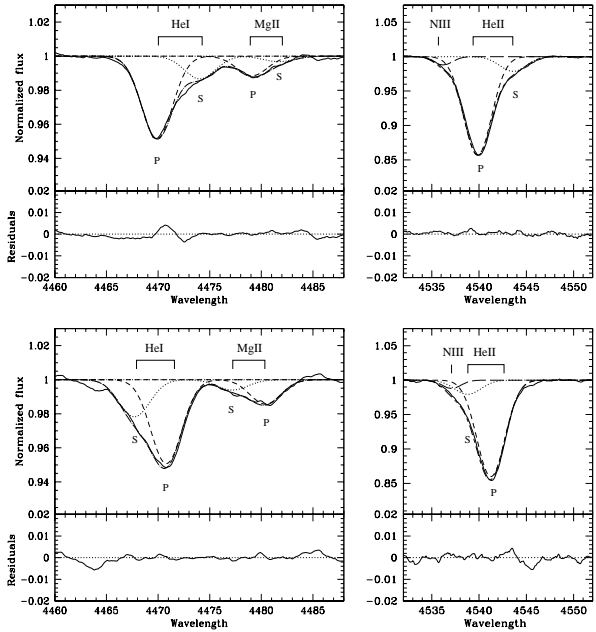}
\caption{\small Disentangling of the line profiles from the two components of HD\,15558 respectively at the minimum of the radial velocity curve ({\it upper part of the figure}), and at the maximum of the radial velocity curve ({\it lower part of the figure}). The two data sets correspond to the mean spectra obtained respectively during the September 2000 and September 2001 observing runs with the 1.52\,m OHP telescope. The profiles of the \ion{He}{i} $\lambda$ 4471, the \ion{Mg}{ii} $\lambda$ 4481, and the \ion{He}{ii} $\lambda$ 4542 lines are displayed, along with the best-fit functions (dotted-dashed lines). The individual Gaussians are also displayed, with the dashed and dotted lines standing respectively for the primary (P) and the secondary (S). The bottom part of each panel represents the residuals (in the sense data -- fit). \label{debl}}
\end{figure}

The upper and lower panels of Fig.\,\ref{debl} show the profile of the \ion{He}{i} $\lambda$ 4471, \ion{Mg}{ii} $\lambda$ 4481 and \ion{He}{ii} $\lambda$ 4542 lines respectively at the minimum and at the maximum of the primary radial velocity curve. We disentangled the profiles from the primary and the secondary by fitting Gaussians following an iterative process. We constructed fitting functions of the form given by Eq.\,\ref{gau}:
\begin{equation}
\label{gau}
P(\lambda)\,=\,\sum_{j}^\mathrm{N}\,\frac{A_{j}}{\sigma_{j}\,\sqrt{2\,\pi}}\,\exp \Big[\,-\,\frac{(\lambda - \lambda_{c,j})^2}{2\,\sigma_{j}^2})\Big]
\end{equation}
with N\,=\,2,3,4 following the complexity of the blended profile. The three parameters to be determined for each Gaussian are respectively the normalization factor ($A_{j}$), the standard deviation of the Gaussian ($\sigma_{j}$), and the central position ($\lambda_{c,j}$). We iteratively changed the values for the three parameters with N\,=\,2 to obtain a first order fit of the primary and of the secondary for the helium lines. We then increased the number of Gaussian components in our fit to account for the presence of the \ion{N}{iii} $\lambda$ 4537 line on the blue side of the \ion{He}{ii} $\lambda$ 4542 line. We optimized the fit of the \ion{He}{ii} line by fitting the \ion{N}{iii} line in the blend at the minimum of the radial velocity curve of the primary, i.e. when the secondary \ion{He}{ii} line is shifted to the red and consequently well separated from the \ion{N}{iii} line. We then fixed the width and the normalization factor of the \ion{N}{iii} line to the same values to fit the helium components of the blended profile at the maximum of the RV curve. We used a N\,=\,4 function to fit the blend of the \ion{He}{i} $\lambda$ 4471 line with that of \ion{Mg}{ii} at 4481\,\AA\, (primary and secondary). The synthetic profiles obtained on the basis of Eq.\,\ref{gau} are overplotted on the observed profiles in Fig\,\ref{debl}. While performing these fits, we first fixed the width and the normalization coefficient of the primary component to the same values at both extrema. We then improved the fits by allowing these parameters to vary.

On the basis of our best fits, we estimated the equivalenth width (EW) of the \ion{He}{i} and \ion{He}{ii} lines to obtain a spectral classification of the two components of the system. At the maximum of the RV curve, we obtain EWs of 0.18 $\pm$ 0.02 and 0.53 $\pm$ 0.03\,\AA\, respectively for the \ion{He}{i} $\lambda$ 4471 and \ion{He}{ii} $\lambda$ 4542 lines of the primary. At the same phase but for the secondary, we obtain respectively EWs of 0.05 $\pm$ 0.02 and 0.08 $\pm$ 0.02\,\AA\,. Using the classification criteria proposed by \citet{Mat}, we obtain O5 and O6.5 spectral types respectively for the primary and the secondary. The best-fit parameters obtained at the phase corresponding to the minimum of the SB1 RV curve lead to EWs of 0.19 $\pm$ 0.02 and 0.52 $\pm$ 0.03\,\AA\, (respectively 0.08 $\pm$ 0.02 and 0.07 $\pm$ 0.03\,\AA\,) for the \ion{He}{i} and \ion{He}{ii} lines of the primary (resp. secondary). In this case, the spectral types of the two components of HD\,15558 are O5.5 and O7.5. Considering mean values of the EWs obtained at the extreme phases of the orbit, the classification of the two components of the system are O5.5 and O7 respectively for the primary and the secondary.

We note that the EW of \ion{He}{i} $\lambda$ 4471 undergoes a significant decrease ($\sim$\,40\%) as the secondary is receding from the observer. Such a decrease is also marginally observed for the \ion{Mg}{ii} line, but no such behaviour is detected for the \ion{He}{ii} $\lambda$ 4542 line. However, the EWs of the primary remain steady from one extremum to the other. This behaviour is quite reminiscent of the so-called Struve-Sahade effect observed in the case of several massive binary system (see e.g. \citealt{Bag}). However, we note that, for such weak lines as in HD\,15558, it might be due to errors in the normalization of the spectra. 

\subsection{Estimation of the masses of the components of HD\,15558 \label{mass}}

As we were able to separate the primary and secondary components from the profiles of a few lines at both extrema of the RV curve, we obtained a first estimate of the amplitude of the secondary RV curve. In the case of the \ion{He}{ii} $\lambda$ 4542 line, we estimated that the RV of the secondary was determined with an uncertainty of about 15\,km\,s$^{-1}$ at its maximum, and about 20\,km\,s$^{-1}$ at its minimum. This difference is due to the fact that the blend with the \ion{N}{iii} line makes the determination of the parameters of the secondary component of the \ion{He}{ii} line more difficult (see bottom right panel of Fig.\,\ref{debl}). According to our fit, and using the $K$ value derived from our SB1 fit (see Table\,\ref{orbpar}), we obtain a mass ratio of 3.3 $\pm$ 0.4. The explicit expression of the mass function as a function of the period ($P$), the eccentricity ($e$), the primary RV semi-amplitude ($K$) and the mass ratio ($q$) allowed us to estimate the minimum masses of the two components. We obtain minimum masses of 152 $\pm$ 51 and 46 $\pm$ 11\,M$_\odot$ respectively for the primary and the secondary. We note that the large errors on the minimum masses are clearly dominated by the large uncertainty on the mass ratio.

The very large masses of the two stars, mainly of the primary, are compatible with the large values already mentioned in the literature for this star (see e.g. \citealt{HPV}). This is the first time that spectroscopic data lend support to this assertion. However, we emphasize that the minimum masses are poorly constrained. The critical issue regarding our results is the estimate of the mass ratio. More RV measurements are needed close to the extrema of the radial velocity curve to reduce the uncertainty on $q$, and accordingly reduce the error bars on the minimum masses. In order to address this issue and to tentatively derive the first SB2 orbital solution for this system, we applied a distentangling method to our spectral time series. 

\subsection{Disentangling of the primary and secondary spectra of HD\,15558\label{disent}}

We used the spectral disentangling method described by \citet{GL}. It consists of an iterative procedure allowing to compute the spectra and RVs of the two components of a binary system. At the starting point, a flat spectrum is adopted for the secondary, and the phase-shifted mean of the observed spectra is used as the starting primary spectrum. At each iteration, the spectra of the primary and the secondary obtained at the previous step are subtracted from the observed spectra, and the radial velocities are determined on the residuals through a cross-correlation procedure using a template including a series of lines. The advantage of this procedure is the fact that it does not require any accurate a priori knowledge of the spectrum of the two stars.

\begin{table*}
\caption{Orbital parameters of the SB2 solution of HD\,15558 determined from the RVs computed with the disentangling procedure of \citet{GL}. We started the iteration procedure with the mean of the \ion{He}{ii} $\lambda$ 4542 and the \ion{N}{iii} $\lambda\lambda$ 4634-4641 lines ({\it left part}) and of the \ion{He}{ii} $\lambda$ 4542 line alone ({\it right part}). The parameters have the same meaning as in Table\,\ref{orbpar}. $R_\mathrm{RL}$ stands for the radius of a sphere with a volume equal to that of the Roche lobe computed according to the formula of \citet{Egg}.\label{paramsb2}}
\begin{center}
\begin{tabular}{l c c c c c}
\hline\hline
 & \multicolumn{2}{c}{\ion{He}{ii} \& \ion{N}{iii}} & & \multicolumn{2}{c}{\ion{He}{ii}}\\
\cline{2-3}\cline{5-6}
 & Primary & Secondary & & Primary & Secondary \\
\hline
P (days)  & \multicolumn{5}{c}{$\sim$\,442 (fixed)} \\
$e$   & \multicolumn{2}{c}{0.41 $\pm$ 0.06} & & \multicolumn{2}{c}{0.37 $\pm$ 0.07}\\
T$_\circ$ (HJD--2\,450\,000) & \multicolumn{2}{c}{1790.377 $\pm$ 11.935} & & \multicolumn{2}{c}{1796.271 $\pm$ 12.970} \\
$\omega$ & \multicolumn{2}{c}{100$^\circ$ $\pm$ 10$^\circ$} & & \multicolumn{2}{c}{120$^\circ$ $\pm$ 12$^\circ$} \\
$\gamma$ (km\,s$^{-1}$)  & --54.1 $\pm$ 8.2 & 0.6 $\pm$ 11.4 & & --51.5 $\pm$ 12.3 & --10.2 $\pm$ 15.3 \\
$K$ (km\,s$^{-1}$) & 48.3 $\pm$ 4.6 & 157.9 $\pm$ 15.2 & & 43.3 $\pm$ 6.3 & 135.9 $\pm$ 19.7 \\
$a\,\sin\,i$ (R$_\odot$) & 385.6 $\pm$ 38.8 & 1259.5 $\pm$ 126.6 & & 351.4 $\pm$ 52.0 & 1101.3 $\pm$ 163.0 \\
$m\,\sin^3\,i$ (M$_\odot$) & 234.0 $\pm$ 61.4 & 71.6 $\pm$ 15.8 & & 159.5 $\pm$ 58.4 & 50.9 $\pm$ 14.4 \\
$q\,=\,m_1/m_2$ & \multicolumn{2}{c}{3.27 $\pm$ 0.45} & & \multicolumn{2}{c}{3.13 $\pm$ 0.71} \\
$R_\mathrm{RL}\,\sin\,i$ (R$_\odot$) & 186.5 $\pm$ 19.3 & 355.8 $\pm$ 37.9 & & 168.6 $\pm$ 26.0 & 314.6 $\pm$ 52.3  \\
\vspace*{-0.2cm}\\
\hline
\end{tabular}
\end{center}
\end{table*}

\begin{figure}
\centering
\includegraphics[width=80mm]{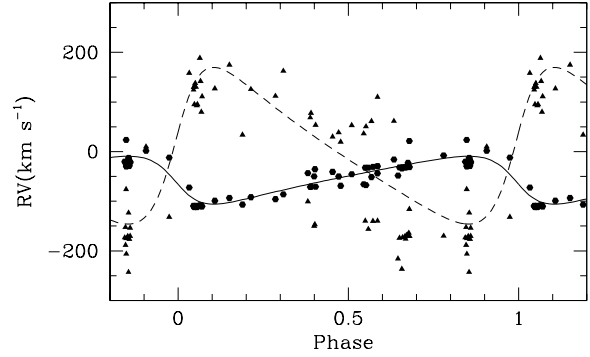}
\caption{Radial velocity curves of the two components of HD\,15558 for an orbital period of $\sim$\,442 d. The hexagons (resp. triangles) stand for the primary (resp. secondary) RVs. The solid and dashed line yield our best fit orbital solution respectively for the primary and the secondary, with the parameters provided in the left part of Table\,\ref{paramsb2}. \label{orbsol2}}
\end{figure}

\begin{figure}
\centering
\includegraphics[width=80mm]{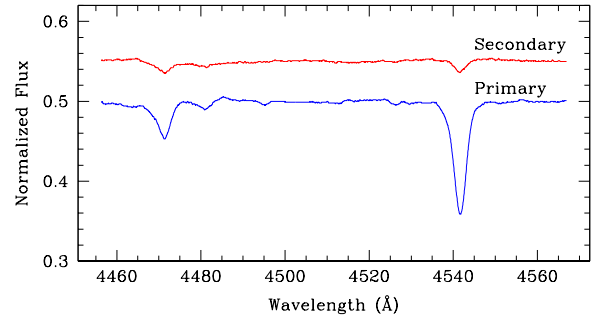}
\caption{Separated spectra of the primary ({\it lower spectrum}) and of the secondary ({\it upper spectrum}) of HD\,15558 in the wavelength domain used for the disentangling procedure. \label{disspec}}
\end{figure}

As our spectral time series include data obtained with different instrumentations, we  once again limited our investigation to a wavelength domain covered by all our data. We applied the disentangling procedure to our 70 spectra between 4456 and 4567\,\AA\,, in order to include the \ion{He}{i} $\lambda$ 4471, the \ion{Mg}{ii} $\lambda$ 4482 and the \ion{He}{ii} $\lambda$ 4542 lines. These lines were selected because they display rather clearly the signature of both components of the system. Starting from the two sets of RVs used to obtain the SB1 solutions described in Table\,\ref{orbpar}, we applied this technique in three different situations:
\begin{enumerate}
\item[(a)] First, we used all our spectra and we allowed both primary and secondary RVs to vary. The RVs obtained after 20 iterations were used to compute an SB2 orbital solution, following the technique described by \citet{sanasb2}. The orbital parameters are given in Table\,\ref{paramsb2}. The most striking results are the huge minimum masses for the stars, mostly for the primary, in agreement with the values discussed in Sect.\,\ref{mass}. We see that most parameters have values close to those obtained for the SB1 orbital solution (see Table\,\ref{orbpar}). Even the difference in the semi-amplitude of the primary radial velocity curve is rather small, whilst the RVs were estimated on the basis of two very different approaches. The Gaussian fit used for the SB1 case likely underestimates the amplitude of the RV curve, but Fig.\,\ref{debl} suggests that the shift between the blended profile and the primary component centroids should be very small. In the case of the mean \ion{He}{ii} and \ion{N}{iii} (resp. \ion{He}{ii} alone) RVs, we note that the disentangling of four (resp. one) spectra gave unsatisfactory results as the derived RVs deviated strongly from those of other spectra obtained only a few days before or after. The disentangling program yields indeed unreasonable radial velocities for the secondary, mostly in the case of spectra with poor signal-to-noise ratios (all deviant points come from Elodie data). For this reason, we discarded these spectra from the computation of the orbital solutions presented in Table\,\ref{paramsb2}. In Fig.\,\ref{orbsol2}, we present the radial velocity curves obtained for both components of the system. We see that the agreement between the data and the computed RV curve for the secondary is rather poor, mostly at phases where the separation between the two components is small. The separated spectra of the primary and of the secondary in the wavelength domain used for the disentangling procedure are individually displayed in Fig.\,\ref{disspec}.
\item[(b)] We then applied the same procedure, but we fixed the radial velocities of the primary to the values taken for the SB1 solution. We obtained rather similar results as in case (a) although the RVs derived for the secondary in the case of several spectra were not acceptable.   
\item[(c)] We finally applied the spectral disentangling method to a reduced spectral series, including only 32 spectra obtained at phases close to the extrema of the radial velocity curve, i.e. between phases 0.05 and 0.25 and between phases 0.75 and 0.95. We selected these spectra in order to obtain an estimate of the mass ratio unaffected by the data obtained when the primary and secondary components are heavily blended. This approach leads to a mass ratio of 3.7 $\pm$ 1.2, but we note that this value should be considered with caution as the eccentricity (0.1 $\pm$ 0.1) and the T$_o$ deviate significantly from the values derived from the SB1 and the SB2 (cases a and b) solutions.
\end{enumerate}

In summary, we find that the mass ratio of the two components of the system obtained from radial velocities determined on the basis of the disentangling method is of the order of 3, whatever the initial RVs considered as the starting point of the procedure. This value is in agreement with the mass ratio derived from the simultaneous fit of Gaussians discussed in Sect.\,\ref{mass}, but we failed to reduce the uncertainty on $q$. Consequently, the minimum masses estimated with both approaches are similar. The extreme minimum masses seem therefore to be robust. For initial RVs obtained with the \ion{He}{ii} $\lambda$ 4542 line, both methods provide minimum masses of the order of 150\,$\pm$\,50 and 50\,$\pm$\,15\,M$_\odot$ respectively for the primary and the secondary. For RVs obtained from the \ion{He}{ii} and \ion{N}{iii} lines, the minimum masses are somewhat larger with a larger uncertainty. Although the uncertainty on the secondary radial velocities is large, the fact that our results rely partly on several spectra obtained close to the extrema of the radial velocity curves lends some support to the mass ratios derived by our analysis.

\section{Discussion} \label{disc}

\subsection{Detailed spectral classification of HD\,15558} \label{class58}
As shown by \citet{thesis}, the \ion{He}{ii} $\lambda$ 4686 line shows a P-Cygni profile for the primary and is in absorption for the secondary. These profiles, along with the luminosity contrast between the primary and the secondary, are compatible with a difference of one luminosity class between the two stars. Considering the fact that H$\alpha$ is in absorption (see Fig.\,\ref{ohp}), it is unlikely that the primary is a supergiant and we therefore propose giant and main-sequence luminosity classes respectively for the primary and the secondary. On the other hand, it is not clear whether the \ion{N}{iii} $\lambda\lambda$ 4634,4641 lines are in emission for the secondary as well. Consequently, we propose that the massive binary HD\,15558 is formed of an O5.5III(f) primary with an O7V secondary.

\subsection{A very massive primary?}

The most interesting -- and also puzzling -- result presented in Sect.\,\ref{hd15558} is the very high minimum mass derived for the primary of HD\,15558, especially considering its spectral type. On the one hand, the most massive star whose mass was determined using spectroscopic and photometric data is the massive binary WR20a, with masses of 82 and 83\,M$_\odot$ respectively for the primary and the secondary (\citealt{wr20a}; \citealt{bonanos}). On the other hand, very high stellar masses were proposed (i) for the Pistol Star, close to the Galactic center, with a mass around 200 - 250\,M$_\odot$ \citep{pistol}, and (ii) for the most massive stars in the Large Magellanic Cloud with masses of the order of 120 - 200\,M$_\odot$ \citep{MH}.

The fact that we derive such a high mass for the primary raises the question of the so-called stellar `upper mass limit'. Massive stars are expected to be vibrationally unstable above a given critical mass which depends on the evolutionary state. For zero-age main sequence stars, the most accurate calculations resulted in critical masses between about 90 and 440 solar masses (see \citealt{appen} for a review). \citet{Oey} statistically demonstrated that the local census of massive stars oberved so far (Milky Way + Magellanic Clouds) exhibits a `universal' upper mass cutoff around 120 - 200\,M$_\odot$ for a Salpeter initial mass function (IMF). Considering that the largest stellar mass observed in IC\,1805 is about 100\,M$_\odot$, the same authors estimate that the probability that the stellar population of IC\,1805 extends to 150\,M$_\odot$ (resp. 200\,M$_\odot$) is about 0.63 (resp. 0.51). As a consequence, a stellar mass of 150\,$\pm$\,50\,M$_\odot$, assuming the minimum masses derived above are close to the true masses ($i\,\sim\,90^\circ$), does not disagree with both the theoretical and the statistical approaches of the stellar upper mass limit.\\  

Provided the mass derived for the primary is indeed of the order of 150\,M$_\odot$\footnote{We rely on the fact that the confidence interval derived by our analysis shows that there is a 65\,\% probability that the minimum mass of the primary lies between about 100 and 200\,M$_\odot$. This does not rule completely out a situation where this mass is lower than 100\,M$_\odot$.}, the main issue comes from its spectral type. We may indeed expect the most massive stars to be also the earliest ones. Considering for instance the typical masses given by \citet{HP}, we see that the expected masses for O5.5 stars should lie between 50 and 80\,M$_\odot$ depending on the luminosity class. An O5.5III star with a mass of about 150\,M$_\odot$ -- i.e. about a factor 2 too high -- constitutes therefore a severe anomaly with respect to the usual classification adopted for O-type stars\footnote{We note also that a mass of 50\,M$_\odot$ for the O7V secondary is somewhat excessive. However, considering the error bar on the mass of the secondary, the 1-$\sigma$ confidence interval is still compatible with the typical values proposed by \citet{HP} for an O7V star, i.e. 36\,M$_\odot$. Consequently, we consider that the mass of the primary is the main issue worth addressing here.}. In addition, a very massive star is also expected to be very luminous. From the observed V magnitude of HD\,15558 and considering a reddening law with R$_V$ = 3.1 and a color excess E(B - V) = 0.71\footnote{We obtained the color excess E(B - V) using the observed (B - V)\,=\,0.5 given by \citet{massey1805} and the intrinsic color (B - V)$_\circ$ based on the spectral type.}, we derive an extinction A$_\mathrm{V}$ equal to 2.20, leading to an apparent dereddened V magnitude equal to 5.70. Using the relation given by \citet{HP}, we obtain a bolometric correction of --3.86, yielding a dereddened apparent bolometric magnitude equal to 1.84. Considering a distance to IC\,1805 of 2.3\,kpc \citep{massey1805}, we derive a log(L$_{bol}$/L$_\odot$) equal to about 5.9. If we consider that the primary contributes about 5/6 of the total bolometric luminosity, log(L$_{bol}$/L$_\odot$) reduces to about 5.82. This value is much too low to be compatible with a 150\,M$_\odot$ primary. Indeed, a confrontation of the bolometric luminosity to theoretical evolutionary tracks for various metallicities \citep{schaller,schaerer} suggests masses not higher than about 70\,M$_\odot$. If the distance we assumed here is correct, such a massive primary should be much more luminous than observed.

On the other hand, an initially extremely massive and very hot early-type (i.e.\ spectral type O2) main sequence star will cool down during its evolution towards the giant luminosity class. For instance, according to the evolutionary models of \citet{schaller}, a star of initial mass 120\,M$_{\odot}$ starts its evolution with an effective temperature of 53000\,K. When the effective temperature reaches 38000\,K, typical for an O5.5 giant \citep{martins}, the stellar mass has decreased to about 85\,M$_{\odot}$. Moreover, at this evolutionary stage, the hydrogen surface abundance should already be reduced, whilst helium should be enhanced. The main conclusion here is that a very massive star would have to lose a substantial fraction of its initial mass before reaching the effective temperature of an O5.5 giant. This would then imply an extremely large initial mass for the primary of HD\,15558.\\

Let us consider an alternative scenario, where HD\,15558 is not a binary but a hierarchical triple system. The primary may be constituted of a yet unrevealed close binary system. In this scenario, as the mass of the primary is estimated on the basis of the motion of the secondary, the primary object -- i.e. the hypothetical close binary -- would appear to be a massive object whose mass is the sum of the masses of two stars. Provided the spectral type derived from our spectra for the primary is typical of the two stars constituting the close binary, we are possibly observing the composite spectrum of two similar O-stars, in addition to that of the secondary object whose spectral type should be O7. This scenario offers the possibility to explain the unexpected high mass of the primary object, and to reconcile its mass with its spectral type. However, the triple system scenario still needs to solve the following issue: our spectral time series did not reveal any binary motion on a time-scale of a few days. If the primary is indeed constituted of two stars, the short period orbit might be seen under a very low inclination angle. We might also consider the possibility that the time-scale of this orbit is a few weeks or so, and therefore poorly sampled by our spectral time series. However, we note that the radial velocity curve plotted in Fig.\,\ref{orbsol} does not present any strong dispersion likely to be due to an orbital motion on a time-scale significantly shorter than the main period of 442\,d. In addition, we might expect some additional X-rays to be produced by an interaction between the winds of the two stars constituting the primary. High quality data obtained for instance with {\it XMM-Newton} are strongly needed to investigate the X-ray emission from HD\,15558 and discuss its origin in detail. In summary, even though our data do not provide any strong evidence supporting the triple scenario, this scenario may reconcile the mass derived for the primary and its spectral type.

\subsection{An open cluster with unusually large stellar masses?}

The existence of such a massive star is worth discussing in the context of the massive star population of IC\,1805, which should not be addressed without considering the presence of the most evolved object known in the vicinity: the microquasar LSI\,+61$^\circ$303. This high mass X-ray binary producing a collimated relativistic jet consists of a Be star and of a compact object whose nature (neutron star or black hole) is not yet established \citep{massilsi61303}. LSI\,+61$^\circ$303 is believed to have been ejected out of IC\,1805 during the supernova explosion of the initially most massive component of the binary \citep{MRL}. If the progenitor of LSI\,+61$^\circ$303 was formed at the same epoch as the other O-stars in IC\,1805, among which is HD\,15558, its primary component may have been more massive than the primary of HD\,15558. Alternatively, if LSI\,+61$^\circ$303 was part of an older population of stars, its supernova explosion might have triggered the formation of the current population of O-type stars.

Two additional stars of this cluster are believed to have very large stellar masses, namely HD\,15570 and HD\,15629 \citep[see][]{HPV}, even though their masses have up to now only been estimated through model atmosphere fits. Unfortunately, as these two stars are probably single, an independant mass determination through the study of a binary motion is unlikely.

This series of presumably very massive objects suggests that IC\,1805 harbours a population of particularly massive stars as compared to other open clusters. According to \citet{massey1805}, a large number of very massive stars  in an open cluster may be explained by its youth. Indeed, the age of the massive star population in IC\,1805 was estimated to be 2\,$\pm$\,1\,Myr. Its most massive members have therefore not yet evolved into compact objects.

\subsection{Multiplicity in IC\,1805} \label{mult}
The multiplicity of massive stars in young open clusters is a crucial question. The \citet{GM1805} proposal of a binary frequency of 80\,\% for O-stars suggests that the star formation process that was at work in IC\,1805 favored the formation of massive binary systems. However, in our spectroscopic study of three mid-O-type stars of this cluster \citep{ic1805_1}, we showed that only one system is a spectroscopic binary (BD\,+60$^\circ$\,497), whilst the other two are most probably single stars (BD\,+60$^\circ$\,501 and BD\,+60$^\circ$\,513), leading to the conclusion that the binary frequency claimed by \cite{GM1805} might be overestimated.

In Table\,\ref{ostars}, we summarize our present knowledge of the multiplicity of the O-star members of IC\,1805. At this stage, spectroscopic monitoring confirmed the binarity of only two members, i.e. HD\,15558 (this paper) and BD\,+60$^\circ$\,497 (Paper I; \citealt{hillwigcasob6}), and our investigations revealed no indication of binarity for four of them (BD\,+60$^\circ$\,501, BD\,+60$^\circ$\,513, HD\,15570 and HD\,15629). For the four remaining O-type members, the multiplicity remains an open question. Consequently, at this stage all we can say is that the O-star binary frequency in IC\,1805 should be of at least 20\,\% but is most probably not more than 60\,\%.  

\begin{table}
\caption{O-type star content of IC\,1805. The first column gives the star number following \cite{vas1805}. For the multiplicity, `s' means that our investigations did not reveal any indication of binarity, and `?' a lack of spectroscopic monitoring. The references for the spectral types are: (1) \citet{massey1805}, (2) Paper I, (3) \citet{und67}, (4) \citet{Ish}, (5) this study. (Notes: $^*$\, the spectral type of the four stars that are not studied in Paper I or in this paper may be uncertain as both O- and B-types were proposed by various authors, $^{**}$\,the spectral type of the secondary should range between O8.5 and O9.5, $^{***}$\,\citet{und67} reported double lines for this star.) \label{ostars}}
\begin{center}
\begin{tabular}{ccccc}
\hline\hline
\# & ID & Sp. Type$^*$ & Ref. & Status \\
\hline
21 &  & O9.5V((f)) & 1 & ? \\
104 & BD\,+60$^\circ$\,497 & O6.5V((f)) & 2 & SB2 \\
 & & + O9V((f))$^{**}$ & & \\
112 & BD\,+60$^\circ$\,498 & O9V & 3 & ?$^{***}$ \\
113 &  & O9Ve & 4 & ? \\
118 & BD\,+60$^\circ$\,499 & O9.5V((f)) & 1 & ? \\
138 & BD\,+60$^\circ$\,501 & O7V((f)) & 2 & s \\
148 & HD\,15558 & O5.5III(f) & 5 & SB2 \\
 & & + O7V & & \\
160 & HD\,15570 & O4If$^+$ & 5 & s \\
192 & HD\,15629 & O5((f)) & 5 & s \\
232 & BD\,+60$^\circ$\,513 & O7.5V((f)) & 2 & s \\
\hline
\end{tabular}
\end{center}
\end{table}

\section{Summary and conclusions\label{conc}} 

We have presented the results of an intensive spectroscopic study of the brightest massive stars in the young open cluster IC\,1805: HD\,15570, HD\,15629 and HD\,15558. For HD\,15570 and HD\,15629, the RVs do not present any significant trend attributable to binary motion on a time scale of a few days, nor from one year to the next. This is in line with the results obtained by \citet{hillwigcasob6} who performed a search for RV variation on time-scales of a few days for some O-type stars of IC\,1805 including HD\,15629 and HD\,15570.

In the case of HD\,15558, we have derived an SB1 orbital solution with significantly refined parameters as compared to those obtained by \citet{GM}. The system appears to be eccentric ($e$\,$\sim$\,0.4) and we obtain a period of 442\,$\pm$\,12\,d. A careful inspection of the spectra obtained close to the extrema of the radial velocity curve reveals the presence of the companion in the profiles of the \ion{He}{i} $\lambda$ 4471, \ion{He}{ii} $\lambda$ 4542, \ion{C}{iv} $\lambda$ 5812 and \ion{He}{i} $\lambda$ 5876 lines. We have simultaneously fitted Gaussians to the profiles of the \ion{He}{i} $\lambda$ 4471 and \ion{He}{ii} $\lambda$ 4542 lines in order to separate the primary and secondary components. The determination of the equivalent width of these two lines allowed us to derive O5.5 and O7 spectral types respectively for both stars. Considering in addition that the \ion{He}{ii} $\lambda$ 4686 line of the primary has a P-Cygni profile whilst the same line is in absorption in the case of the secondary, along with the fact that we do not clearly observe the \ion{N}{iii} $\lambda$$\lambda$ 4634-4641 lines in emission for the secondary, we propose that HD\,15558 is an O5.5III(f) + O7V binary.\\

We estimated the radial velocities of the secondary of HD\,15558 following two approaches: (1) a simultaneous fit of line profiles and (2) a disentangling method. Both techniques allowed us to determine minimum masses of the order of 150\,$\pm$\,50 and 50\,$\pm$\,15\,M$_\odot$ respectively for the primary and the secondary. We also obtained the first SB2 orbital solution for HD\,15558. Although we note that the quality of this SB2 solution is rather poor, our data point clearly to a rather high mass ratio (about 3), leading to an extreme minimum mass for the primary. Our results require independent validation using an improved disentangling procedure.

The main problem in considering our results is to reconcile the very extreme mass of the primary with its spectral type. It is indeed unlikely that a very massive main-sequence star could cool down enough during its evolution to become an O5.5 giant. A possible scenario can however be considered where HD\,15558 is not a binary but a triple system. The primary may be a yet unrevealed close binary system. In this case, as the mass of the primary is estimated on the basis of the motion of the secondary, the primary object -- i.e.\,the hypothetical close binary -- would appear to be a massive object whose mass is the sum of the masses of two stars. Even though our data do not provide any evidence for this scenario, we estimate that at this stage it should not be rejected, and that it could constitute a valuable working hypothesis for future investigations concerning this system.

From our new and previously published results, we briefly address the question of the multiplicity of the early-type stars in IC\,1805. The binary frequency among O-stars should be of at least 20\,\%, since out of 10 O-stars only 2 are confirmed binaries, and should not exceed 60\,\%. We therefore conclude that the previously claimed binary frequency of 80\,\% was overestimated.

\begin{acknowledgements}
We gratefully acknowledge the anonymous referee for the careful reading, and for comments that significantly helped us to improve the paper. We are indebted to the FNRS (Belgium) for assistance including contract 1.5.051.00 "Cr\'edit aux chercheurs". The travels to OHP were supported by the Minist\`ere de l'Enseignement Sup\'erieur et de la Recherche de la Communaut\'e Fran\c{c}aise. This research is also supported in part by contract PAI P5/36 (Belgian Federal Science Policy Office) and through the PRODEX XMM/Integral contract. We thank the staff of the Observatoire de Haute Provence (France) and of the San Pedro Martir Observatory (Mexico) for their technical support. Part of our OHP data were obtained in Service Mode. We are grateful to all the observers and technicians who took good care of our observations. The SIMBAD database has been consulted for the bibliography.
\end{acknowledgements}

%\bibliographystyle{aa}
%\bibliography{bibmd}

\Online

\appendix
\section{Observations and data reduction}\label{obs}
Spectroscopic observations were collected at the Observatoire de Haute-Provence (OHP, France) during several observing runs from 2000 to 2004 for HD\,15570 and HD\,15558, and from 2002 to 2004 for HD\,15629. All spectra were obtained with the Aur\'elie spectrograph fed by the 1.52\,m telescope \citep{aurelie}, using the same setup as described in Paper I. Our collection of spectra is described in Table\,\ref{runs}.

\begin{table}
\caption{Observing runs used for the line profile variability study of HD\,15570, HD\,15629 and HD\,15558. The first and second columns give the name of the campaign as used in the text as well as the instrumentation used. The next columns are the number of spectra obtained, the time elapsed between the first and the last spectrum of the run, the natural width of a peak of the power spectrum taken as 1/$\Delta$T, and the mean signal-to-noise ratio of each data set.\label{runs}}
\begin{center}
\begin{tabular}{l c c c c c}
\hline\hline
 Obs. run & Telescope & N & $\Delta$T & $\Delta\nu_\mathrm{nat}$ & S/N \\
  &  &  & (d) & (d$^{-1}$) & \\
\hline
\multicolumn{2}{l}{HD\,15570} & & & & \\
\hline
\vspace*{-0.2cm}\\
Sept. 2000 & OHP/1.52\,m & 7 & 9.99 & 0.10 & 390 \\
Sept. 2001 & OHP/1.52\,m & 5 & 7.02 & 0.14 & 360 \\
Sept. 2002 & OHP/1.52\,m & 10 & 14.94 & 0.07 & 300 \\ 
Oct. 2003 & OHP/1.52\,m & 7 & 17.97 & 0.06 & 300 \\
Oct. 2004 & OHP/1.52\,m & 7 & 9.90 & 0.10 & 260 \\
Oct. 2004 & SPM/2.10\,m & 4 & 2.07 & 0.48 & 180 \\
\vspace*{-0.2cm}\\
\hline
\multicolumn{2}{l}{HD\,15629} & & & & \\
\hline
\vspace*{-0.2cm}\\
Sept. 2002 & OHP/1.52\,m & 9 & 12.99 & 0.08 & 300 \\
Oct. 2003 & OHP/1.52\,m & 4 & 17.98 & 0.06 & 280 \\
Oct. 2004 & OHP/1.52\,m & 7 & 9.90 & 0.10 & 260 \\
\vspace*{-0.2cm}\\
\hline
\multicolumn{2}{l}{HD\,15558} & & & & \\
\hline
\vspace*{-0.2cm}\\
Sept. 2000 & OHP/1.52\,m & 12 & 10.97 & 0.09 & 530 \\
Sept. 2001 & OHP/1.52\,m & 12 & 7.02 & 0.14 & 440 \\
Sept. 2002 & OHP/1.52\,m & 11 & 14.92 & 0.07 & 430 \\
Oct. 2003 & OHP/1.52\,m & 7 & 17.95 & 0.06 & 360 \\
Oct. 2004 & OHP/1.52\,m & 5 & 7.00 & 0.14 & 450 \\
\vspace*{-0.2cm}\\
\hline
\end{tabular}
\end{center}
\end{table}

We also obtained several spectra of HD\,15570 with the echelle spectrograph mounted on the 2.1\,m telescope at the Observatorio Astron\'omico Nacional of San Pedro Martir (SPM) in Mexico, with exposure times ranging from 10 to 20 minutes. The instrument covers the spectral domain between about 3800 and 6800\,\AA\,. The detector was a Site CCD with 1024\,$\times$\,1024 pixels of 24\,$\mu$m$^2$. The slit width was set to 200\,$\mu$m corresponding to 2\,arcsec on the sky. The data were reduced using the echelle package available within the MIDAS software. After adding consecutive spectra of a given night to reach higher signal-to-noise ratios, at the expense of time resolution, we obtained 4 spectra of HD\,15570.\\ 

We obtained 17 additional spectra of HD\,15558 with the Elodie echelle spectrograph \citep{elodie} fed by the 1.93\,m telescope at the Observatoire de Haute Provence between March 2003 and February 2005 to monitor a complete orbital period. The exposure time of each of these spectra was 90 minutes. This spectrograph uses a combination of a prism and a grism as a cross-disperser, with a blaze angle of 76$^\circ$. The resolving power achieved is about 42000 between 3906 and 6811\,\AA\, in a single exposure, and the detector is a Tk1024 CCD with 24\,$\mu$m\,$\times$\,24\,$\mu$m pixels. The Elodie data consist of single spectra distributed over 67 orders. Due to pointing constraints specific to the 1.93\,m telescope, no echelle spectra were obtained between April 2004 and July 2004. We filled that gap with 6 observations with the Aur\'elie spectrograph mounted on the 1.52\,m telescope. Two spectra out of the 6 were obtained using the same grating as described in the previous paragraph (between 4455 and 4890\,\AA\,). The four remaining spectra were obtained using a 1200 l/mm grating providing a resolving power of about 16000 in the blue range, with a reciproqual dispersion of 8 \AA\,mm$^{-1}$ (between 4455 and 4680\,\AA\,). The exposure time of these 6 Aur\'elie spectra was 60 minutes.

For Aur\'elie data, we adopted the same reduction procedure as in Paper I.  Elodie data were reduced using the standard on-line automatic treatment implemented at the OHP. All spectra were normalized using splines calculated on the basis of properly chosen continuum windows.

\begin{table}[ht]
\caption{Description of the data of HD\,15558 obtained during the long-term monitoring using the 1.93 and the 1.52\,m telescopes. The first column gives the date of the observation. The instrumentation used to obtain the spectrum is provided in the second column. The next columns give the resolving power of the instrumentation used, and the signal-to-noise ratio.\label{monit}}
\begin{center}
\begin{tabular}{l c c c}
\hline\hline
\vspace*{-0.2cm}\\
 Date & Telescope/Instr. & $\lambda/\Delta\lambda$ &  S/N \\
  &  &  &  \\
\vspace*{-0.2cm}\\
\hline
\vspace*{-0.2cm}\\
2003/09/04 & 1.93\,m/Elodie & 42000 & 170 \\
2003/10/04 & 1.93\,m/Elodie & 42000 & 160 \\
2003/10/21 & 1.93\,m/Elodie & 42000 & 100 \\
2004/01/07 & 1.93\,m/Elodie & 42000 & 180 \\
2004/01/15 & 1.93\,m/Elodie & 42000 & 100 \\
2004/02/14 & 1.93\,m/Elodie & 42000 & 130 \\
2004/03/11 & 1.93\,m/Elodie & 42000 & 110 \\
2004/04/04 & 1.93\,m/Elodie & 42000 & 110 \\
2004/04/10 & 1.52\,m/Aur\'elie & 8000 & 330 \\
2004/05/06 & 1.52\,m/Aur\'elie & 16000 & 160 \\
2004/06/09 & 1.52\,m/Aur\'elie & 16000 & 290 \\
2004/06/28 & 1.52\,m/Aur\'elie & 16000 & 360 \\
2004/07/15 & 1.52\,m/Aur\'elie & 8000 & 550 \\
2004/07/26 & 1.52\,m/Aur\'elie & 16000 & 360 \\
2004/08/15 & 1.93\,m/Elodie & 42000 & 100 \\
2004/08/26 & 1.93\,m/Elodie & 42000 & 120 \\
2004/09/06 & 1.93\,m/Elodie & 42000 & 140 \\
2004/10/07 & 1.93\,m/Elodie & 42000 & 130 \\
2004/11/09 & 1.93\,m/Elodie & 42000 & 60 \\
2004/11/17 & 1.93\,m/Elodie & 42000 & 140 \\
2004/12/03 & 1.93\,m/Elodie & 42000 & 140 \\
2005/01/27 & 1.93\,m/Elodie & 42000 & 100 \\
2005/02/16 & 1.93\,m/Elodie & 42000 & 70 \\
\vspace*{-0.2cm}\\
\hline
\vspace*{-0.2cm}\\
\end{tabular}
\end{center}
\end{table}

\section{Radial velocity measurements}
The mean RVs of HD\,15570 and HD\,15629 obtained for the various observing runs are collected in Table\,\ref{rvtab}. In the case of HD\,15558, all the RVs used to obtain the orbital solution are listed in Table\,\ref{rvtab2}. All the RVs considered in this paper were estimated using the rest wavelengths provided by \citet{CLL}.

\begin{table}
\caption{Radial velocity of the main He and N lines for both stars measured on OHP spectra (expressed in km\,s$^{-1}$). The mean radial velocity along with its 1-$\sigma$ standard deviations are provided for each observing run. All individual RVs can be found in the WEBDA data base at {\tt http://obswww.unige.ch/webda}. \label{rvtab}}
\begin{center}
\tiny
\begin{tabular}{c c c c c c}
\hline
 & \multicolumn{4}{c}{HD\,15570} \\
\cline{2-5}
Data set & \ion{He}{i} & \ion{He}{ii} & \ion{N}{iii} & \ion{N}{iii} \\
 & $\lambda$ 4471 & $\lambda$ 4542 & $\lambda$ 4634 & $\lambda$ 4641 \\
\hline
Sept.2000 & --69.1 $\pm$ 8.2 & --46.7 $\pm$ 8.7 & --56.1 $\pm$ 3.9 & --67.0 $\pm$ 2.8 \\
Sept.2001 & --65.1 $\pm$ 10.6 & --51.1 $\pm$ 3.2 & --58.6 $\pm$ 3.5 & --70.3 $\pm$ 2.0 \\
Sept.2002 & --60.5 $\pm$ 13.2 & --50.0 $\pm$ 5.0 & --56.8 $\pm$ 4.6 & --70.0 $\pm$ 3.3 \\
Oct.2003 & --59.3 $\pm$ 15.7 & --46.2 $\pm$ 3.9 & --55.6 $\pm$ 4.5 & --67.6 $\pm$ 5.4 \\
Oct.2004 & --72.7 $\pm$ 8.3 & --47.8 $\pm$ 7.0 & --58.0 $\pm$ 4.7 & --70.6 $\pm$ 3.7 \\
All & --65.0 $\pm$ 12.3 & --48.3 $\pm$ 5.9 & --56.9 $\pm$ 4.2 & --69.1 $\pm$ 3.8 \\
\hline
 & \multicolumn{4}{c}{HD\,15629} \\
\cline{2-5}
Data set & \ion{He}{i} & \ion{He}{ii} & \ion{N}{iii} & \ion{N}{iii} \\
 & $\lambda$ 4471 & $\lambda$ 4542 & $\lambda$ 4634 & $\lambda$ 4641 \\
\hline
Sept.2000 & -- & -- & -- & -- \\
Sept.2001 & -- & -- & -- & -- \\
Sept.2002 & --60.6 $\pm$ 3.6 & --46.9 $\pm$ 4.5 & --59.9 $\pm$ 8.8 & --72.1 $\pm$ 6.8 \\
Oct.2003 & --55.3 $\pm$ 4.6 & --41.0 $\pm$ 1.6 & --54.6 $\pm$ 3.2 & --71.6 $\pm$ 2.6 \\
Oct.2004 & --58.2 $\pm$ 5.9 & --50.1 $\pm$ 4.6 & --63.5 $\pm$ 8.3 & --74.6 $\pm$ 6.5 \\
All & --58.7 $\pm$ 4.9 & --46.8 $\pm$ 5.2 & --60.1 $\pm$ 8.2 & --72.9 $\pm$ 6.0 \\
\hline
\end{tabular}
\end{center}
\end{table}
\normalsize

\begin{table}[h]
\caption{Radial velocities obtained from our time series of HD\,15558. The second and third columns give the heliocentric Julian day (--2\,450\,000) and the orbital phase following the parameters provided in Table\,\ref{orbpar} (left part). The next columns provide the radial velocities obtained for lines that were selected for the determination of the SB1 orbital solution.  The column labelled `Mean' contains the mean of the radial velocities obtained for the \ion{He}{ii} and the \ion{N}{iii} lines quoted in the fourth and fifth columns. All RVs are expressed in km\,s$^{-1}$. The last column gives the weight (W) attributed to our measurements to calculate the orbital parameters, depending on the spectral resolution of the instrument and the signal-to-noise ratio of the spectra. \label{rvtab2}}
\begin{center}
\begin{tabular}{l c c c c c c }
\hline\hline
\vspace*{-0.2cm}\\
\# & HJD & $\phi$ & \ion{He}{ii} & \ion{N}{iii} & Mean &  W \\
 &  &  & $\lambda$\,4542 & $\lambda$$\lambda$\,4634, &  & \\
 &  &  &  & 4641 &  & \\
\hline
\vspace*{-0.2cm}\\
1 & 1810.640 & 0.034 & --85.9 & --96.1 & --91.0 & 0.25\\
2 & 1810.658 & 0.034 & --88.0 & --117.0 & --102.5 & 0.25\\
3 & 1811.619 & 0.037 & --86.7 & --109.0 & --97.9 & 0.25\\
4 & 1811.665 & 0.038 & --88.8 & --101.5 & --95.1 & 0.25\\
5 & 1812.660 & 0.039 & --87.5 & --112.6 & --100.0 & 0.25\\
6 & 1813.651 & 0.041 & --87.8 & --118.8 & --103.3 & 0.25\\
7 & 1814.643 & 0.043 & --91.6 & --118.9 & --105.2 & 0.25\\
8 & 1815.645 & 0.046 & --88.9 & --110.2 & --99.6 & 0.25\\
9 & 1818.641 & 0.053 & --86.4 & --110.6 & --98.5 & 0.25\\
10 & 1819.593 & 0.055 & --86.8 & --109.9 & --98.3 & 0.25\\
11 & 1820.648 & 0.057 & --83.0 & --118.1 & --100.6 & 0.25\\
12 & 1821.609 & 0.059 & --91.2 & --111.2 & --101.2 & 0.25\\
13 & 2163.580 & 0.832 & --1.2 & --12.1 & --6.7 & 0.25\\
14 & 2164.597 & 0.834 & --4.3 & --10.4 & --7.3 & 0.25\\
15 & 2164.660 & 0.835 & --0.9 & --18.8 & --9.9 & 0.25\\
16 & 2165.577 & 0.837 & --3.5 & --23.7 & --13.6 & 0.25\\
17 & 2165.590 & 0.837 & --0.1 & --12.9 & --6.5 & 0.25\\
18 & 2167.550 & 0.841 & --1.9 & --7.4 & --4.6 & 0.25\\
19 & 2167.564 & 0.841 & --6.7 & --14.8 & --10.7 & 0.25\\
20 & 2168.601 & 0.843 & --1.8 & --17.3 & --9.5 & 0.25\\
21 & 2168.642 & 0.844 & --6.0 & --18.9 & --12.4 & 0.25\\
22 & 2169.581 & 0.846 & --4.2 & --16.8 & --10.5 & 0.25\\
23 & 2170.580 & 0.848 & --7.2 & --22.1 & --14.6 & 0.25\\
24 & 2170.603 & 0.848 & --8.1 & --23.9 & --16.0 & 0.25\\
25 & 2518.617 & 0.634 & --27.5 & --28.5 & --28.0 & 0.25\\
26 & 2520.593 & 0.639 & --14.8 & --20.2 & --17.5 & 0.25\\
27 & 2523.580 & 0.646 & --11.5 & --32.8 & --22.2 & 0.25\\
28 & 2524.510 & 0.648 & --13.8 & --28.0 & --20.9 & 0.25\\
29 & 2527.531 & 0.655 & --12.4 & --21.6 & --17.0 & 0.25\\
30 & 2528.499 & 0.657 & --14.5 & --26.7 & --20.6 & 0.25\\
31 & 2529.529 & 0.659 & --10.5 & --21.6 & --16.1 & 0.25\\
32 & 2531.503 & 0.664 & --4.9 & --21.8 & --13.4 & 0.25\\
33 & 2532.516 & 0.666 & --8.3 & --20.6 & --14.5 & 0.25\\
34 & 2532.644 & 0.666 & --7.9 & --27.2 & --17.5 & 0.25\\
35 & 2533.532 & 0.668 & --12.6 & --24.2 & --18.4 & 0.25\\
36 & 2886.627 & 0.466 & --37.0 & --51.5 & --44.2 & 1.00\\
37 & 2916.592 & 0.534 & --29.4 & --62.0 & --45.7 & 1.00\\
38 & 2916.558 & 0.534 & --24.7 & --39.1 & --31.9 & 0.25\\
39 & 2918.570 & 0.538 & --25.7 & --52.6 & --39.2 & 0.25\\
40 & 2919.561 & 0.541 & --26.6 & --47.8 & --37.2 & 0.25\\
41 & 2922.566 & 0.547 & --24.4 & --34.4 & --29.4 & 0.25\\
42 & 2926.573 & 0.556 & --19.6 & --34.8 & --27.2 & 0.25\\
43 & 2928.593 & 0.561 & --24.7 & --39.4 & --32.0 & 0.25\\
44 & 2934.511 & 0.574 & --20.0 & --33.2 & --26.6 & 0.25\\
\vspace*{-0.2cm}\\
\hline
\end{tabular}
\end{center}
\end{table}
\normalsize

\setcounter{table}{1}

\begin{table}[ht]
\caption{(continued)}
\begin{center}
\begin{tabular}{l c c c c c c}
\hline\hline
\vspace*{-0.2cm}\\
\# & HJD & $\phi$ & \ion{He}{ii} & \ion{N}{iii} & Mean & W \\
 &  &  & $\lambda$\,4542 & $\lambda$$\lambda$\,4634, &  & \\
 &  &  &  & 4641 &  & \\
\hline
\vspace*{-0.2cm}\\
45 & 2934.481 & 0.574 & --13.7 & --47.1 & --30.4 & 1.00\\
46 & 3012.476 & 0.751 & --14.0 & --31.5 & --22.7 & 1.00\\
47 & 3020.329 & 0.768 & --19.2 & --20.7 & --20.0 & 1.00\\
48 & 3050.375 & 0.836 & --11.2 & --24.6 & --17.9 & 1.00\\
49 & 3076.317 & 0.895 & --16.7 & --34.3 & --25.5 & 1.00\\
50 & 3100.318 & 0.949 & --43.1 & --62.1 & --52.6 & 1.00\\
51 & 3106.340 & 0.963 & --28.8 & --47.1 & --38.0 & 0.25\\
52 & 3132.374 & 0.022 & --63.5 & --85.2 & --74.3 & 0.50\\
53 & 3165.550 & 0.097 & --88.2 & --114.4 & --101.3 & 0.50\\
54 & 3184.590 & 0.140 & --82.7 & --104.4 & --93.6 & 0.50\\
55 & 3201.593 & 0.178 & --72.1 & --98.7 & --85.4 & 0.25\\
56 & 3212.544 & 0.203 & --65.3 & --94.9 & --80.1 & 0.50\\
57 & 3232.579 & 0.248 & --67.1 & --87.5 & --77.3 & 1.00\\
58 & 3244.384 & 0.275 & --63.0 & --93.0 & --78.0 & 1.00\\
59 & 3254.630 & 0.298 & --46.3 & --78.0 & --62.2 & 1.00\\
60 & 3286.477 & 0.370 & --43.8 & --60.1 & --52.0 & 1.00\\
61 & 3289.570 & 0.377 & --46.3 & --62.9 & --54.6 & 0.25\\
62 & 3290.553 & 0.379 & --35.6 & --53.3 & --44.4 & 0.25\\
63 & 3294.603 & 0.388 & --32.3 & --46.6 & --39.4 & 0.25\\
64 & 3295.584 & 0.390 & --37.6 & --44.2 & --40.9 & 0.25\\
65 & 3296.569 & 0.393 & --44.0 & --56.0 & --50.0 & 0.25\\
66 & 3318.587 & 0.442 & --28.5 & --41.8 & --35.1 & 0.10\\
67 & 3326.508 & 0.460 & --33.4 & --56.5 & --45.0 & 1.00\\
68 & 3343.487 & 0.499 & --27.9 & --54.4 & --41.2 & 1.00\\
69 & 3398.308 & 0.623 & --9.4 & --33.8 & --21.6 & 1.00\\
70 & 3418.295 & 0.668 & --3.6 & --24.5 & --14.0 & 0.10\\
\vspace*{-0.2cm}\\
\hline
\end{tabular}
\end{center}
\end{table}

\end{document}